\documentclass[11pt]{article}
\usepackage{epsf}

\newcommand\reference{\smallskip\par\noindent\hangindent 15pt}

\title{The Center is Everywhere}

\author{David H. Weinberg\thanks{Department of Astronomy, 
Ohio State University, Columbus, OH 43210}}
\date{September 23, 2012}

\topmargin = -\topskip
\advance\topmargin by -\headsep
\begin{document}
\maketitle

\begin{abstract}
{\it The Center is Everywhere} is a sculpture by Josiah McElheny, 
currently (through October 14, 2012) on exhibit at the Institute of 
Contemporary Art, Boston. The sculpture is based on data from the Sloan Digital
Sky Survey (SDSS), using hundreds of glass crystals and lamps suspended from 
brass rods to represent the three-dimensional structure mapped by the SDSS
through one of its 2000+ spectroscopic plugplates. This article describes
the scientific ideas behind this sculpture, emphasizing the 
principle of the statistical homogeneity of cosmic structure in the
presence of local complexity.  The title of the sculpture is inspired by 
the work of the French revolutionary Louis Auguste Blanqui, whose 1872 book 
{\it L'\'Eternit\'e
par les astres: Hypoth\`ese astronomique} was the first to raise the spectre
of the infinite replicas expected in an infinite, statistically homogeneous
universe.  Puzzles of infinities, probabilities, and replicas 
continue to haunt modern fiction 
and contemporary discussions of inflationary cosmology. 
\end{abstract}

\section{A Principle, A Map, A Sculpture}

As beings bound to a small planet orbiting an
unexceptional star in the outskirts of a galaxy that is
one among trillions, how can we hope to understand the
universe?  Astronomers approach this problem by drawing on two key ideas.
The first is the Cosmological Principle, an infinite extrapolation
of the Copernican discovery that displaced the earth from
its special location at the center of everything.
In technical terms, the Cosmological Principle states that the universe
on large scales is homogeneous and isotropic: it has no preferred
places and no preferred directions.  If the Cosmological Principle
holds, then the same laws of physics apply everywhere, and when we observe
a region of the universe large enough to be statistically
representative we can extrapolate its properties to the rest
of the cosmos.  The second idea is the finite speed of light.
When we observe a distant object, we see it as it was when
the light left it.  This fact turns telescopes into time machines ---
from our vantage point in the present, we can peer back into
cosmic history.
Josiah McElheny's sculpture {\it The Center is Everywhere} is
based on observations from the Sloan Digital Sky Survey (SDSS),
which has created the largest ever
maps of the distant universe.\footnote{From this planet, anyway.}

In detail, the Cosmological Principle is manifestly wrong.
It is different to look up than to look down, different to look
to the center of the Milky Way galaxy than to look outward
from our location in its stellar disk, different to look at nearby galaxies
like the Magellanic Clouds or the Andromeda Nebula than to 
look elsewhere in the sky.
But when we use telescopes to
reach far beyond our local neighborhood, the Cosmological Principle holds
remarkably well.  While stars are collected in galaxies
and galaxies themselves cluster in groups and filaments interweaved
with tunnels of emptiness, the average properties of 
galaxies and clusters and filaments are similar in different
regions of the universe.  Furthermore, when we observe the
most distant source of ``light'' that we can see, a background of microwaves
that has been traveling freely since the universe became
transparent 300,000 years after the Big Bang, we see that it is 
astonishingly smooth, implying that the distribution of matter at that
time was uniform to within a tiny fraction of a percent.  
Gravity has since amplified primordial fluctuations, the tiny imperfections 
of that infant universe, into galaxies and larger structures,
but the fluctuations themselves were random, so every
place in the universe initially
had the same chance to become ``interesting'' or ``boring.''

Over the past decade and a half, the SDSS has used a dedicated
telescope at Apache Point Observatory in New Mexico to obtain digital
images over 1/4 of the sky and to map the 3-dimensional distribution
of more than 2 million galaxies, 1 million stars, and 200,000 quasars.
The stars are all members of the Milky Way at distances of hundreds
or thousands of light years; their separation in time from us is less than
a cosmic eyeblink.  The galaxies, each a collection of tens or hundreds
of billions of stars, span a wide range of distances, from thousands
of light years to billions of light years.  Quasars are the most
luminous objects in the universe, powered by hot gas falling onto
supermassive black holes, so they can be seen to very large distances.
The most distant quasars discovered by the SDSS have lookback times
of more than 13 billion years, and the light we detect left them less
than a billion years after the Big Bang.

To map millions of galaxies, the SDSS observes them using
``fiber plugplates,'' each an aluminum disk $33"$ in diameter drilled
with 640 holes at the locations of objects selected for observation
from the digital images.  Each hole is plugged with an optical
fiber, and all 640 fibers are fed back into spectrographs riding on the
back of the telescope, which disperse the light into its constituent
colors and allow astronomers to infer the properties and distance
of each object from subtle patterns of dark and bright lines, which are 
produced when atoms absorb and emit light at particular wavelengths.
{\it The Center is Everywhere} utilizes a reproduction of one of the
SDSS plugplates, number 1945, replacing each optical fiber with
a brass suspension rod whose terminating piece represents
the astronomical object that the SDSS observed through that hole on this
plugplate: small glass spheres for stars in the Milky Way,
lamps for the quasars, glass disks for the flattened, spinning galaxies
like the Milky Way, and larger spheres for the 
rounded ``elliptical'' galaxies that form when gravity drags
disk galaxies together and scrambles their stars onto disordered orbits.
The length of each rod corresponds to the light-travel distance to
the object, with the longest rod in the sculpture representing a 
light ray that has traveled 12.3 billion years to reach 
the earth.\footnote{For the astronomers in the audience: the mapping
from distance to rod length changes from linear at small 
distances to logarithmic at large distances, following
$l = 72" \times \ln(1+t/4\,{\rm Gyr})/\ln(1+t_0/4\,{\rm Gyr})$,
where $t$ is the lookback time and $t_0$ is present age of
the universe computed for a cosmology with $\Omega_m=0.25$,
$\Omega_\Lambda=0.75$, and $H_0 = 70\,{\rm km}\,{\rm s}^{-1}\,{\rm Mpc}^{-1}$.}

Disk galaxies are more common than ellipticals, but the ellipticals
are, on average, bigger and brighter, so the SDSS telescope
can see them to greater distances.  Scanning downwards from the
top of {\it The Center is Everywhere}, a dense pattern of disks
gradually gives way to a sparser sprinkling of ellipticals.  
Quasars are much rarer still, for while most large galaxies contain
supermassive black holes, it is only rare, short-lived events
that funnel gas toward them and cause them to flare into
brilliance.  At the peak of its activity, a quasar can outshine
its galactic host by a factor of 1000, so quasars can be detected
to the farthest reaches of the universe.
The volume mapped by a single plugplate expands like the cone of
a lighthouse beam, so at large enough distances it begins to encompass
many quasars, rare though they are.
Stars enter the SDSS in many ways: some are observed as standard
sources to calibrate measurements of galaxies and quasars, some are
mapped in a deliberate pattern to reveal the structure of the
Milky Way, and some are ``contaminants,'' stars with unusual colors
that allow them to masquerade as quasars until an SDSS spectrum reveals
their true identity.

\begin{figure}
\begin{centering}
\epsfysize=12.3truecm
\epsfbox{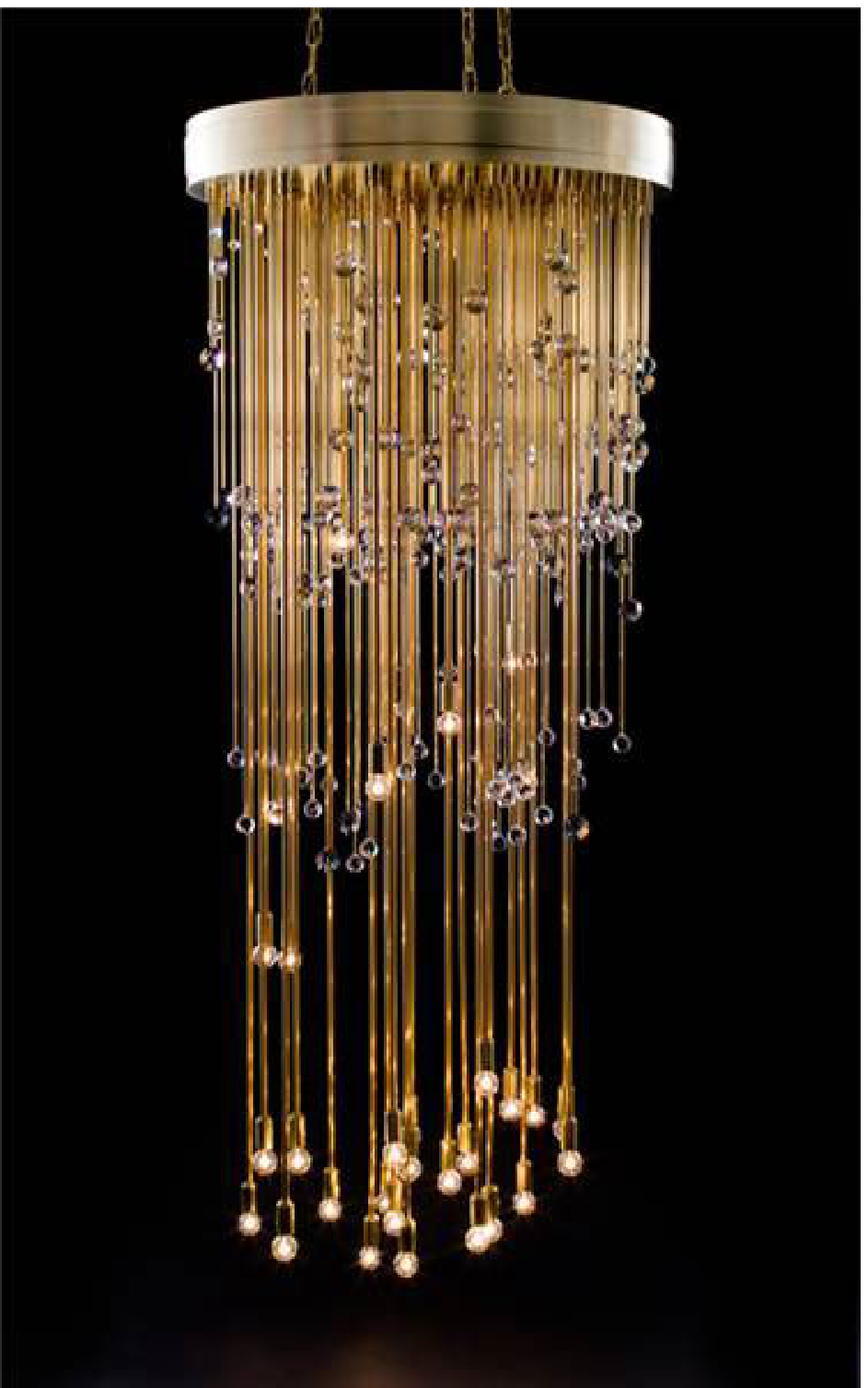}
\hskip 0.15truein
\epsfysize=12.3truecm
\epsfbox{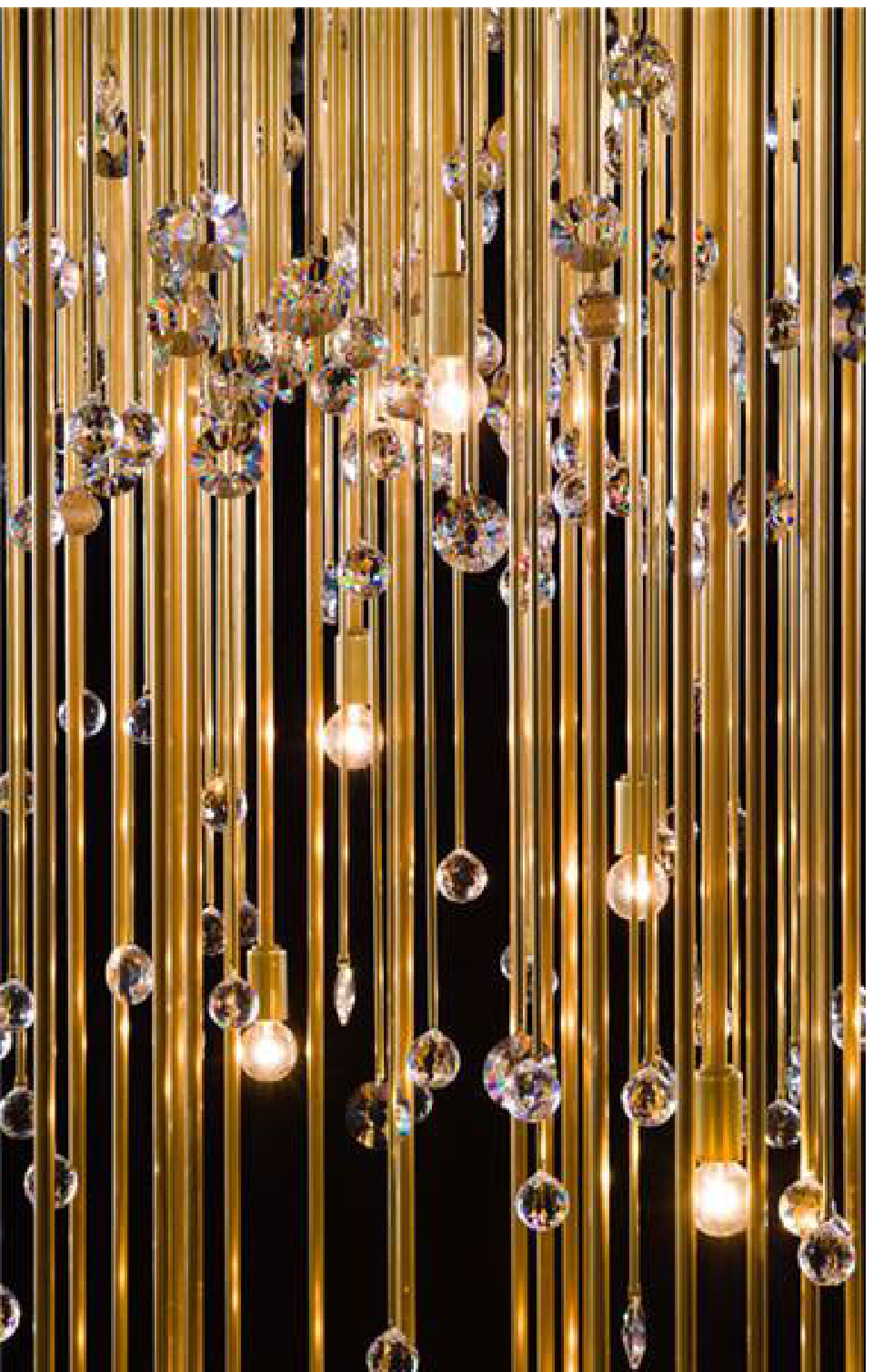}
\end{centering}
\caption{Full view (left) and detail (right) of
{\it Study for The Center is Everywhere},'' by
Josiah McElheny (2012), designed in collaboration with David Weinberg.
The sculpture is 32 inches in diameter and 84 inches high, made of brass, 
steel, three forms of cut lead crystals (representing stars, 
disk galaxies, and elliptical galaxies), and electric bulbs 
(representing quasars).  The structure is based on spectroscopic plugplate 
1945 from the Sloan Digital Sky Survey.
}
\end{figure}

Even a casual inspection of {\it The Center is Everywhere} shows that
galaxies are not randomly distributed through space.
The gravity that pulls gas and dark matter into galaxies
also pulls the galaxies into groups and clusters, arranged on
still larger scales in a web of filaments and walls.
A defining goal of the SDSS is to measure the clustering
of galaxies with exquisite precision and detail, 
measurements that help to 
reveal the history of galaxy formation, pin down the geometry,
matter, and energy content of the cosmos, and probe the physics
of the Big Bang itself.  
The cosmic core sample shown in {\it The Center is Everywhere}
is one small chunk of the SDSS map of the universe, a random
but representative piece of the whole.
An individual SDSS plugplate covers an area on the sky that is 
three degrees across, six times the diameter
of the full moon, the size of a quarter held at arm's length.
The circular plates must overlap so that they do not leave gaps between them,
and in its original phase (see Appendix A)
the SDSS used more than 2000 plates to
map 1/4 of the sky, one quarter-sized patch at a time.

An astute viewer with the patience to count will notice that
{\it The Center is Everywhere} has 306 rods, not the expected 640.
The minimum separation of holes in the original SDSS plugplate 
is only $0.08"$, an uncomfortably small size to impose on the 
cut-glass crystals of a chandelier.  {\it The Center is Everywhere}
circumvents this problem with three compromises: it expands the central
region of plate 1945 and trims its edges, it slightly nudges
the positions of some holes and the lengths of some rods to
create space, and it omits some objects that still can't fit.

The full title of McElheny's sculpture is {\it Study for The Center
is Everywhere}.  It is a ``study'' in part because this piece is
the first of what may well become a series, with the multitude
of SDSS plugplates providing an ample library for sculptures
that follow common rules but are individually unique.  But even
a series of 50 such sculptures would still be ``a study'' for the
much larger work that could be constructed in principle, with
2000+ chandeliers suspended in an arena-like vault, holding a million
glass crystals and lamps to represent the celestial objects that
have been mapped by the SDSS.  And even this vast conceptual installation
would remain ``a study'': the complete work should take its
title seriously and include the maps constructed by astronomers
and artists in other, distant galaxies, whose own 
Studies for {\it The Center is Everywhere} include the Milky Way
as a glimmering disk of glass.

\section{The Unsettling Infinite}

{\it ``Nature is an infinite sphere, whose center is
everywhere and whose circumference is nowhere.''}  The French philosopher
Blaise Pascal wrote this description ({\it Pens\'ees, 1669})
centuries before the emergence of the modern cosmological model,
or the gravitational theory of Einstein on which it is based.
Yet it seems an apt encapsulation of the Cosmological Principle, 
of the centerless expansion of the Big Bang theory,
and of our consequent
ability to map the universe fairly from an arbitrary point within it.
Because the universe has a finite age --- about 14 billion years 
according to recent observations --- the size of the {\it observable}
universe is, in fact, finite.  A galaxy or quasar more than
14 billion light years away is beyond the edge of our cosmic
horizon, and it will remain so forever.  But just as the slight
curvature of the earthly horizon viewed from a ship bespeaks a world far larger
than the sailors can see, so astronomical observations (which,
despite increasingly precise measurements, have detected {\it no}
curvature of 3-dimensional space)
tell us that the universe is at least hundreds 
of times larger than the span of our cosmic
horizon, probably much larger, 
and quite possibly infinite.  If the Cosmological Principle
holds across an infinite universe, then there are other astronomers
mapping their own observable sphere so far from us that we could 
never communicate, not even in principle.

Pascal's metaphor has a long history, as described by Jorge Luis Borges
in his essay {\it The Fearful Sphere of Pascal}.  One who took it
up was the communist revolutionary Louis Auguste Blanqui, who devoted
his imprisonment to writing a cosmological treatise,
{\it Eternity Through The Stars: An Astronomical Hypothesis} (1872).
On the first page of his book, Blanqui repeats and slightly amends
Pascal's comment: ``{\it The universe is a sphere whose center
is everywhere and whose surface is nowhere.}''
Blanqui becomes the first to clearly state one of the most bizarre
properties of an infinite universe made of constituent atoms
and molecules whose variety and possible arrangements are finite.
In such a universe, every chance event, and every configuration
of matter that can arise from a series of such events, must occur,
and not just once, but an infinite number of times.  If we take the idea
of the infinite seriously, then mathematical reasoning leads us to
the dizzying notion that the (very) distant universe holds replicas of
ourselves, reading the same books, speaking the same words, thinking
the same thoughts . . .  and near replicas who do the same thing but in a
different language, or with different colored eyes.  Blanqui's thought
experiment is like that of the Librarians in the Borges story
{\it The Library of Babel}, who conjecture that the Library is
infinite and conclude that its volumes of seemingly random letters
must somewhere include all the works of great literature, and
of bad literature, in all variations, over, and over, and over.

Computations of probabilities in an infinite universe present
conceptual and philosophical challenges even today,
forming an esoteric and sometimes controversial thread
of discussion among contemporary cosmologists and string
theorists.  The stakes of this discussion are real, for it tells
us whether an underlying fundamental theory can, with reasonable
probability, explain the world as we see it.
Observations do not tell us whether the universe is truly
infinite, nor whether it remains statistically homogeneous
far beyond our horizon, so there are avenues of escape from
the replicas that Blanqui envisioned.  But the mere possibility
that we inhabit an infinite universe stocked with infinite
replicas disturbs our already troubled senses of individuality
and free will.

Closer to home, our finite observable universe is already enormous,
and even the SDSS has mapped only a tiny fraction of its contents.
Space is big and time is long, and these vastnesses can be intimidating.
But one can choose instead to marvel at the particularities of
nature, at the intricacy of its structures,
at the way that gravity and electromagnetism and nuclear forces
weave random fluctuations of primordial atoms into a web of galaxies, stars,
and planets.  On some of these planets, chemistry and biology
and evolution organize those atoms into beings who can, from their own center
of the universe, map the cosmos and begin to understand it.

\bigskip
\noindent{\bf Acknowledgment}

\medskip\noindent
I gratefully acknowledge support from the National Science Foundation,
via grant AST-1009505.  I thank Josiah McElheny
for permission to reproduce images of 
{\it Study for The Center is Everywhere}, and for
the opportunity to participate in a remarkable collaboration.

\vfill\eject
\appendix
\section{About the Sloan Digital Sky Survey}

The SDSS was conceived around 1990, and it began operation in 1998
after nearly a decade of design and construction.
In its initial phase, the SDSS used the world's largest digital camera
and fiber-fed spectrographs, both mounted on a dedicated 2.5-meter telescope,
to obtain deep multi-color images of about 20\% of the sky and spectra
of nearly a million stars, galaxies, and quasars.  In its second
phase (SDSS-II, 2005-2008), the Sloan collaboration completed the
project's original goals and carried out two additional surveys,
one a map of stars in the Milky Way and one a systematic search
for supernovae to measure the history of cosmic expansion.
In its current phase (SDSS-III, 2008-2014), the collaboration is
using the same telescope equipped with upgraded and new spectroscopic
instruments to conduct four surveys, of the distant universe, the
Milky Way galaxy, and extra-solar planetary systems.  The team is already
far along in planning for expanded surveys beyond 2014.
Technical descriptions of the SDSS include York et al.\ (2000) for
SDSS-I, Abazajian et al.\ (2008) for SDSS-II, and
Eisenstein et al.\ (2011) for SDSS-III.

The ``Sloan'' in SDSS refers to the Alfred P. Sloan Foundation, which 
has provided critical funding support to all phases of the SDSS.
The SDSS has also been supported by the participating institutions,
by the U.S. National Science Foundation, Department of Energy 
Office of Science, and National Aeronautics and Space Agency,
and by funding agencies of other nations.  Today the SDSS collaboration
has more than 500 active scientists from 40+ institutions around
the globe.  More information about the SDSS can be found at the
web sites {\tt sdss.org} and {\tt sdss3.org}.  The public archive of
SDSS images and spectra is available through {\tt skyserver.sdss3.org}.

\section{About Josiah McElheny and David Weinberg}

Josiah McElheny is an artist who works primarily in glass, 
though his recent {\it oeuvre} includes films, books, and other media.
Much of his work is created in his glass-blowing studio in Brooklyn, NY.
His intellectual interests are wide ranging and include the rise and fall
of modernist design and architecture, the history of glass as a
material and as an artistic medium, conceptions of the infinite, 
and contemporary cosmology.  Among other honors, he was awarded a 2006
MacArthur Foundation Fellowship.

David Weinberg is a cosmologist and a Professor of Astronomy at
Ohio State University.  He studies the formation and clustering of galaxies,
the structure of the intergalactic medium,
and the matter and energy contents of the universe.
He was the Spokesperson for SDSS-II, and he is the Project Scientist
of SDSS-III.

McElheny and Weinberg have collaborated since 2004 on the design
of several cosmologically inspired sculptures.  
The article by Weinberg (2010)
describes the scientific and intellectual background to the first four of 
these collaborative efforts: {\it An End to Modernity} (2005, now in the
collection of London's Tate Modern gallery), 
{\it The Last Scattering Surface} (2006, now in the Phoenix Art Museum),
{\it The End of the Dark Ages} (2008, now in a private collection),
and {\it Island Universe} (2008).  The 2012 exhibition 
{\it Josiah McElheny: Some Pictures of the Infinite}, at the 
Boston Institute of Contemporary Art, includes both {\it Island Universe}
(previously exhibited in London and Madrid) and the first exhibition
of {\it Study for The Center is Everywhere}.
Further description of this collaboration can be found in the article by
Weinberg (2011).

\vfill\eject
\centerline{\bf BIBLIOGRAPHY}
\bigskip

\reference
Abazajian, K.,
Adelman-McCarthy, J.,
Agueros, M. A.,
Allam, S. S.,
Allende Prieto, C.,
An, D.,
et al., 2009,
``The Seventh Data Release of the Sloan Digital Sky Survey,''
{\it The Astrophysical Journal Supplement Series}, 182, 543.

\reference
Blanqui, L.\ A., 1872, {\it L'\'Eternit\'e par les astres:
Hypoth\`ese astronomique}, first published in Paris by 
Librairie Germer Balli\`ere and printed by E.\ Martinet.
English translation by D.\ Canty, edited by J. McElheny,
published in 2012 by the Donald Young Gallery (Chicago).

\reference
Borges, J. L., ``The Fearful Sphere of Pascal,'' 
in {\it Labyrinths}, ed. D.\ A.\ Yates \& J. E.\ Irby,
published in 1962 by New Directions (New York).

\reference
Borges, J. L., ``The Library of Babel,'' 
in {\it Labyrinths}, ed. D.\ A.\ Yates \& J. E.\ Irby,
published in 1962 by New Directions (New York).

\reference
Eisenstein, D. J.,
Weinberg, D. H.,
Agol, E.,
Aihara, H.,
Allende Prieto, C.,
Anderson, S. F.,
et al., 2011,
``SDSS-III: Massive Spectroscopic Surveys of the Distant Universe,
the Milky Way Galaxy, and Extra-Solar Planetary Systems,''
{\it The Astronomical Journal}, 142, 72.

\reference
Molesworth, H. (editor), 2012, 
{\it Josiah McElheny: Some Pictures of the Infinite}, 
published by The Institute of Contemporary Art, 
Boston (Hatje Cantz Press).

\reference
Weinberg, D.\ H., 2010,
``From the Big Bang to the Multiverse: Translations in Space and Time,''
in {\it Josiah McElheny: A Prism}, eds. L. Neri \&
J. McElheny, Skira/Rizzoli Books (New York).
Available as arXiv:1006.1012.

\reference
Weinberg, D.\ H., 2011,
``From the Big Bang to {\it Island Universe}: Anatomy of a Collaboration,''
{\it Narrative}, 19, 258.
Available as arXiv:1006.1013.

\reference
York, D. G.,
Adelman, J.,
Anderson, J. E.,
Anderson, S. F.,
Annis, J.,
Bahcall, N. A., et al., 2000,
``The Sloan Digital Sky Survey: Technical Summary,''
{\it The Astronomical Journal}, 120, 1579.

\end{document}